\documentclass[12pt]{article}

\newcommand{\be}{\begin{equation}}
\newcommand{\ee}{\end{equation}}
\newcommand{\bi}[1]{\vspace{-3mm} \bibitem{#1}}

\usepackage{epsfig,amsmath,amssymb,graphics,graphicx} 
\usepackage{amscd}
\usepackage{pb-diagram}
\usepackage{hyperref}
\usepackage{amsfonts}

\begin{document}

\begin{center}

{\it International Journal of Statistical Mechanics.
Vol.2014. (2014) 873529.}
\vskip 3mm

{\bf \large Fractional Diffusion Equations for Lattice and Continuum:
\vskip 2mm
Gr\"unwald-Letnikov Differences and Derivatives Approach} \\

\vskip 7mm
{\bf \large Vasily E. Tarasov} \\
\vskip 3mm

{\it Skobeltsyn Institute of Nuclear Physics,\\ 
Lomonosov Moscow State University, \\
Moscow 119991, Russia} \\
{E-mail: tarasov@theory.sinp.msu.ru} \\

\begin{abstract}
Fractional diffusion equations 
for three-dimensional lattice models 
based on fractional-order differences of 
the Gr\"unwald-Letnikov type are suggested. 
These lattice fractional diffusion equations
contain difference operators that describe 
long-range jumps from one lattice site to other.
In continuum limit, the suggested lattice
diffusion equations with non-integer order differences
give the diffusion equations with
the Gr\"unwald-Letnikov fractional derivatives
for continuum.
We propose a consistent derivation of 
the fractional diffusion equation 
with the fractional derivatives of
Gr\"unwald-Letnikov type.
The suggested lattice diffusion equations 
can be considered 
as a new microstructural basis 
of space-fractional diffusion in nonlocal media.
\end{abstract}

\end{center}




\section{Introduction}

Fractional calculus and differential equation
of non-integer orders 
\cite{SKM1,SKM2,KST,Mainardi1997,FC2} 
have a long history that is connected with 
the names of famous scientists such as
Liouville, Riemann, Gr\"unwald, Letnikov, Marchaud and Riesz.
Derivatives and integrals of non-integer orders 
have a lot of application in different areas of physics 
\cite{Mainardi,TarasovSpringer,KLM,MS2012,US}.
Fractional calculus is a powerful tool
to describe processes in
continuously distributed media with nonlocal properties.
As it was shown in \cite{JPA2006,JMP2006}, 
the continuum equations with fractional derivatives 
are directly connected \cite{TarasovSpringer}
to lattice models with long-range interactions. 
The lattice equations for fractional nonlocal media 
and the correspondent continuum equations 
have been considered recently in 
\cite{CEJP2013,ISRN-CMP2014,IJSS2014}.
Fractional-order differences and
the correspondent derivatives 
have been first proposed by Gr\"unwald \cite{Grunwald} 
and by Letnikov \cite{Letnikov}.
At the present time these generalized 
differences and derivatives are called 
the Gr\"unwald-Letnikov fractional differences
and derivatives \cite{SKM1,SKM2,KST,Podlub}.
One-dimensional lattice models
with long-range interactions of the Gr\"unwald-Letnikov type
and the correspondent fractional differential and 
integral continuum equations 
have been suggested in \cite{MOM2014}.
The suggested form of long-range interaction
is based on the form of 
the left-sided and right-sided Gr\"unwald-Letnikov 
fractional differences.
A possible form of lattice vectors calculus based on
the fractional-order differences of the Gr\"unwald-Letnikov type
has been suggested in \cite{JPA2014}.
In this paper, we apply this approach
to describe diffusion on lattices with long-range jumps
and to derive fractional diffusion equations for
nonlocal continuum with power-law nonlocality.


The diffusion equations describe the change of 
probability of a random function in space and time
in transport processes, and it usually have the form of
second-order partial differential equation of parabolic type.
Unfortunately for complex nonlocal media,
the usual second-order diffusion equation 
cannot adequate describe real processes. 
For example, the diffusion processes
with the Poissonian waiting time and 
the L\'evy distribution for the jump length
cannot be described by equation with 
second-order derivatives with respect to coordinates.
The L\'evy distribution describes the L\'evy flights 
\cite{BG1990,MK2000} that are random walks, 
where the jump lengths
have probability distributions with heavy-tails.
The L\'evy motion can be described by equation 
with spatial derivatives of non-integer orders $\mu$, 
\cite{MK2000}. 
In this case, the moment of order $\delta$ 
for the L\'evy motion has the form
$\langle |x(t)|^{\delta} \rangle \sim t^{\delta / \mu}$,
where $0< \delta <\mu \le 2$.
Usually the space-fractional diffusion equations
are obtained from the second-order differential equations
by replacing the second-order space derivatives by 
fractional-order derivatives. 
Fractional diffusion equations with coordinate 
derivatives of non-integer order 
have been suggested in \cite{Zaslavsky1994}. 
The solutions and properties of these
equations are considered in \cite{SZ1997,MN2002,Zaslavsky2002}.
The diffusion equations with fractional coordinate derivatives
was also considered in \cite{KLM,MS2012,Physica2008}.

In this paper we propose a consistent derivation of
the space-fractional diffusion equation 
with Gr\"unwald-Letnikov derivatives of non-integer orders
by using lattice models with long-range jumps 
that is considered as new microstructural basis
to describe fractional diffusion processes in nonlocal media.
In this paper, we suggest a lattice equation 
for probability density of particle 
in unbounded homogeneous three-dimensional lattice 
with long-range jumps between lattice sites. 
We prove that continuous limit for
the suggested lattice diffusion equation gives 
the space-fractional diffusion equation 
for non-local continuum.
The fractional differential equation for continuum contains 
the Gr\"unwald-Letnikov derivatives on non-integer orders.
Continuum models can be considered 
as a continuous limit of lattice models,
where the length-scales of a continuum element 
are much larger than the distances between the lattice sites. 
The suggested approach to derive 
the space fractional diffusion equations 
can serve as a microstructural basis to
describe the spatial-fractional diffusion processes.


\section{Fractional diffusion equation for lattice}

The differences of fractional order and
the correspondent fractional derivatives have been 
introduced by Gr\"unwald in 1867 and independently 
by Letnikov in 1868. 

Differences of non-integer orders are defined as 
a generalization of the integer-order difference 
\be \label{DFO}
\nabla^{m}_{a, \pm} f(x) = 
\sum^{m}_{n=1} \frac{(-1)^n\, m!}{n! \, (m-n)!} \,
f(x \mp n \,a) , \quad 
(a \in \mathbb{R}_{+}, \quad m \in \mathbb{N}) .
\ee
The Fourier transformation {\cal F} 
of the fractional-order differences of Gr\"unwald-Letnikov type
(see Section 20 of \cite{SKM1,SKM2}) has the form
\be \label{FT-FD}
{\cal F} \left\{ \nabla^{\alpha}_{a, \pm} f(x) \right\} (k) = 
\left(1 - e^{\pm \, a \, k} \right)^{\alpha} \,
{\cal F} \left\{ f \right\} (k)  .
\ee
The differences of integer orders (\ref{DFO}) are defined 
by the finite series.
The differences of non-integer order $\alpha \in \mathbb{R}_{+}$ 
are defined as infinite series (see Section 20 of \cite{SKM1,SKM2}).
Fractional-order differences of Gr\"unwald-Letnikov type
are defined by the equation
\be \label{FOD}
\nabla^{\alpha}_{a, \pm} f(x) = \sum_{n = 0}^{\infty} 
\frac{(-1)^n \, \Gamma(\alpha+1)}{\Gamma(n+1) 
\Gamma (\alpha-n+1)} \,f (x \mp n a) , \quad (a>0) .
\ee
The difference $\nabla^{\alpha}_{a, +}$ 
is called left-sided fractional difference, 
and $\nabla^{\alpha}_{a, -}$ 
is called the right-sided fractional difference.
We note that the series in (\ref{FOD}) 
converges absolutely and uniformly 
for every bounded function $f(x)$ and $\alpha>0$. 

Using the fractional-order differences (\ref{FOD}),
we can consider the derivatives of non-integer orders.
The left- and right-sided Gr\"unwald-Letnikov derivatives 
of fractional order $\alpha>0$ are defined by the equation
\be \label{GLD-0a}
^{GL}D^{\alpha}_{x, \pm} f(x) = \lim_{a \to 0+} 
\frac{\nabla^{\alpha}_{a, \pm} f(x)}{|a|^{\alpha}} .
\ee
For integer values of $\alpha = m \in \mathbb{N}$ 
the Gr\"unwald-Letnikov derivatives (\ref{GLD-0a})
are equal to the usual integer-order
derivatives up to the sign in the form
\be \label{GL-integer}
^{GL}D^{m}_{x, \pm} f(x) = ( \pm 1)^m \, 
\frac{d^m f(x)}{d x^m} . 
\ee
We can note that the Gr\"unwald-Letnikov fractional derivatives 
coincide with the Marchaud fractional derivatives 
for the functions from $L_r(\mathbb{R})$, 
where $1 \leqslant r <\infty$ 
(see Theorem 20.4 in \cite{SKM1,SKM2}).


Let us consider three-dimensional unbounded 
and bounded lattices.
Physical lattices are characterized by space periodicity.
For unbounded lattices 
we can use three non-coplanar vectors 
${\bf a}_1$, ${\bf a}_2$, ${\bf a}_2$, that 
are the shortest vectors by which a lattice can 
be displaced and be brought back into itself. 
Sites of this lattice can be characterized by 
the number vector ${\bf n} = (n_1,n_2,n_3)$, 
where $n_j$ ($j=1,2,3$) are integer. 
For simplification, we consider a lattice with mutually 
perpendicular primitive lattice vectors
${\bf a}_j$, ($j=1,2,3$).
We choose directions of the axes of the Cartesian coordinate system coincide with the vector ${\bf a}_j$. 
In this case ${\bf a}_j = a_j \, {\bf e}_j$, 
where $a_j=|{\bf a}_j|>0$ and ${\bf e}_j$ are 
the basis vectors of the Cartesian coordinate system.
This means that we use a primitive orthorhombic Bravais lattice.
Then the vector ${\bf n}$ can be represented as
${\bf n}=n_1{\bf e}_1+n_2{\bf e}_2+n_3{\bf e}_3$. 

Choosing a coordinate origin at one of the lattice sites, 
then the positions of all other site 
with ${\bf n} = (n_1,n_2,n_3)$ is described by the vector 
${\bf r}({\bf n})=n_1{\bf a}_1+n_2{\bf a}_2+n_3{\bf a}_3$. 
The lattice sites are numbered by ${\bf n}$, so 
that the vector ${\bf n}$ can be considered as a 
number vector of the corresponding particle. 
We assume that the positions of particles 
in the lattice coincide with the lattice sites.
The distribution function, which describes
probability density for lattice site ${\bf n}$,
will be denoted by $f ({\bf n},t) = f (n_1,n_2,n_3,t)$.
This function satisfies the conditions
\be
\sum^{+\infty}_{ n_1 = -\infty } \sum^{+\infty}_{ n_2 = -\infty } 
\sum^{+\infty}_{ n_3 = -\infty } f (n_1,n_2,n_3,t) = 1 ,
\quad f (n_1,n_2,n_3,t) \ge 0 \quad (t \ge 0) .
\ee

To describe dynamics of the distribution function
$f ({\bf n},t)$ in the lattice models with long-range jumps
between sites, we define fractional-order 
difference operators 
of the Gr\"unwald-Letnikov type 
in the direction ${\bf e}_j={\bf a}_j/|{\bf a}_j|$
of the lattice.
Fractional-order difference operators 
of the Gr\"unwald-Letnikov type 
for unbounded lattice are the operators 
$^{GL}\mathbb{D}^{\pm}_L \left[ \alpha_j \atop j \right]$
that act on the function $f({\bf m},t)$ as
\be \label{Ds-GL} 
^{GL}\mathbb{D}^{\pm}_L \left[ \alpha_j \atop j \right] \, 
f({\bf m},t) 
= \frac{1}{a^{\alpha}_j} \sum_{m_j=-\infty }^{+\infty} \,
^{GL}K^{\pm}_{\alpha_j} (n_j - m_j) \,
f ({\bf m},t) \quad (\alpha_j >0 , \quad j=1,2,3) ,
\ee
where the kernels $\, ^{GL}K^{\pm}_{\alpha} (n)$ 
are defined by the equation
\be \label{GL-K}
^{GL}K^{\pm}_{\alpha_j} (n) = 
\frac{(-1)^{n} \, \Gamma (1+\alpha_j)\, (H[n] \pm H[-n])}{2 \, \Gamma (|n|+1) \,\Gamma (1+\alpha_j - |n| ) } , 
\quad (\alpha_j >0 , \quad n \in \mathbb{Z}),
\ee
and $\Gamma(z)$ is the gamma function, 
$H[n]$ is the discrete variable Heaviside step function
that is defined as
$H[n] =1$ for $n \ge 0$, and $H[n] =0$ for $n< 0$,
where $n \in \mathbb{Z}$.
The parameter $\alpha_j$ is called the order of the operator.
It should be notes that the definition of $H[0]=1$ for
discrete variable Heaviside function is significant
for us, since it allows us to write the kernels 
$^{GL}K^{+}_{\alpha} (n)$ 
in the simple form without allocating repeated zero terms.
Fractional-order difference operators 
(\ref{Ds-GL}) can be called a lattice 
fractional partial derivative 
in the direction ${\bf e}_j={\bf a}_j/|{\bf a}_j|$.

It should be noted that one-dimensional lattice models 
with the long-range interaction of 
the form $\, ^{GL}K^{+}_{\alpha}(n)$ 
and correspondent fractional nonlocal continuum models
have been suggested in \cite{MOM2014}.
The lattice operators (\ref{Ds-GL}) recently have been proposed
in \cite{JPA2014}.

It is easy to see that the kernels 
$\, ^{GL}K^{+}_{\alpha}(n)$ and $\, ^{GL}K^{-}_{\alpha}(n)$ 
are even and odd functions such that
$\, ^{GL}K^{\pm}_{\alpha}(-n) = 
\pm \, ^{GL}K^{\pm}_{\alpha}(n)$. 
The form of the lattice operators (\ref{Ds-GL}) 
can be defined by the addition and subtraction
of the Gr\"unwald-Letnikov fractional differences 
\be \label{Ds-GL-sum} 
^{GL}\mathbb{D}^{\pm}_L \left[ \alpha \atop j \right] \, 
f({\bf m},t) =
\frac{1}{|a_j|^{\alpha}} \sum_{m_j = 0}^{\infty} 
\frac{(-1)^{m_j} \, \Gamma(1+\alpha)}{2 \, \Gamma(m_j+1) 
\Gamma (1+\alpha-m_j)} \, 
\Bigl( f({\bf n} - m_j {\bf e}_j,t) \pm 
f({\bf n} + m_j {\bf e}_j,t) \Bigr) .
\ee
We should note that in equation (\ref{Ds-GL-sum})
the summation is realized over non-negative values $m_j$, 
in contrast to the sum over all integer values 
in equation (\ref{Ds-GL}). 

For bounded physical lattice models 
the fractional-order difference operators also can be defined. 
Fractional-order difference operators 
of the Gr\"unwald-Letnikov type 
for bounded lattice with $m^1_j \le m_j \le m^2_j$ 
are the operators 
$_B^{GL}\mathbb{D}^{\pm}_L \left[ \alpha_j \atop j \right]$
that act on the function $f({\bf m},t)$ as
\be \label{Ds-GL-B} 
_B^{GL}\mathbb{D}^{\pm}_L \left[ \alpha_j \atop j \right] \, 
f({\bf m},t) = \frac{1}{a^{\alpha}_j} \sum_{m_j=m^1_j}^{m^2_j} \,
^{GL}K^{\pm}_{\alpha_j} (n_j - m_j) \,
f ({\bf m},t) \quad (j=1,2,3) ,
\ee
where the kernels $\, ^{GL}K^{\pm}_{\alpha_j} (n)$ 
are defined by equations (\ref{GL-K}).
The suggested forms of fractional difference operators 
for bounded physical lattice models are based on
the Gr\"unwald-Letnikov fractional differences 
on finite intervals (see Section 20.4 of \cite{SKM1,SKM2}).
For the finite interval $[x^1_j,x^2_j]$, the integer values
$m^1_j$, $m^2_j$ and $m_j$ are defined by the equations 
\be \label{M1M2}
m^1_j = \left[ \frac{x^1_j}{a_j} \right] , \quad
m^2_j = \left[ \frac{x^2_j}{a_j} \right] , \quad
m_j = \left[ \frac{x_j}{a_j} \right] ,
\ee
where the brackets $[ \ \ ]$ of (\ref{M1M2}) mean 
the floor function that maps a real number to 
the largest previous integer number.


Using the semigroup property 
for fractional differences of non-negative orders
(see Property 2.29 in \cite{KST}), 
it is easy to show that the semi-group property holds
for the fractional operators (\ref{Ds-GL}) in the form
\be \label{SGP-LFD}
^{GL}\mathbb{D}^{\pm}_L \left[ \alpha \atop j \right] \, 
^{GL}\mathbb{D}^{\pm}_L \left[ \beta \atop j \right] \, 
 = \,
^{GL}\mathbb{D}^{\pm}_L \left[ \alpha +\beta \atop j \right] \, 
 , \quad (\alpha>0 , \quad \beta>0) .
\ee
Using this equation, it is easily to prove the commutativity 
and the associativity of the lattice operator 
(\ref{Ds-GL}) of the Gr\"unwald-Letnikov type.
The commutativity and associativity of 
the fractional operators (\ref{Ds-GL}) 
of the Gr\"unwald-Letnikov type for different directions
are obvious.

To describe isotropic physical lattices
we should use the difference operators
$^{GL}\mathbb{D}^{\pm}_L \left[ \alpha_j \atop j \right]$ and
$_B^{GL}\mathbb{D}^{\pm}_L \left[ \alpha_j \atop j \right]$ 
with orders $\alpha_j=\alpha$ for all $j=1,2,3$.


Let us give possible equations for distribution 
function $f ({\bf n},t)$ on unbounded and bounded lattices. 
For unbounded homogeneous lattice
the diffusion equation for probability density 
can be considered in the form
\be \label{LattEq-2}
\frac{\partial f ({\bf n},t) }{\partial t} =
- \sum^3_{i=1} g_i \,
\, ^{GL}\mathbb{D}^{\pm} \left[ \alpha_i \atop i \right]
\, f ({\bf m} ,t) ,
+ \sum^3_{i,j=1} g_{ij} \, ^{GL}\mathbb{D}^{\pm,\pm} 
\left[ \alpha_i \, \beta_j \atop i \ j \right] 
 \, f ({\bf m} ,t) .
\ee
For bounded lattice we should use
the fractional difference operators (\ref{Ds-GL-B}), 
and the correspondent analog of equation (\ref{LattEq-2})
has the form
\be \label{LattEq-3}
\frac{\partial f ({\bf n},t)}{\partial t} =
- \sum^3_{i=1} g_i \,
\, ^{GL}_B\mathbb{D}^{\pm} \left[ \alpha_i \atop i \right]
\, f ({\bf m} ,t) ,
+ \sum^3_{i,j=1} g_{ij} \, ^{GL}_B\mathbb{D}^{\pm,\pm} 
\left[ \alpha_i \, \beta_j \atop i \ j \right] 
 \, f ({\bf m} ,t) .
\ee
Equations (\ref{LattEq-2}) and (\ref{LattEq-3}) 
are the three-dimensional 
lattice diffusion equations that describe 
fractional diffusion processes with the lattice jumps.
Here $f({\bf n},t)$ is the probability density function
to find the test particle at site ${\bf n}$ at time $t$.
The italics $i,j \in \{1;2;3\}$ are the coordinate indices, 
$g_{i}$ and $g_{ij}$ are lattice coupling constants. 

The first and second terms of the right hand side of 
equations (\ref{LattEq-2}) and (\ref{LattEq-3}) describe
the particle drift and diffusion on the lattice.
These correspondent kernels describe the long-range drift and 
diffusion to ${\bf n}$-site from all other ${\bf m}$-sites. 
The parameters $\alpha_i$ and $\beta_j$ 
in the kernels are positive real numbers 
that characterize how quickly the intensity of 
the drift and diffusion processes in the lattice
decrease with increasing the value ${\bf n} - {\bf m}$.
The kernels $K^{\pm}_{\alpha_j}(n_j-m_j)$, where $j=1,2,3$, 
describe long-range jumps in the direction ${\bf a}_j$
with lattice step length $|n_j-m_j|$ in the lattice.
In equation (\ref{LattEq-2}), we use the combination of 
the lattice operators
\be \label{Dsss1a}
^{GL}\mathbb{D}^{\pm,\pm} \left[ \alpha_i \, \beta_j \atop i \ j \right] =
^{GL}\mathbb{D}^{\pm} \left[ \alpha_i \atop i \right] \,
^{GL}\mathbb{D}^{\pm} \left[ \beta_j \atop j \right] ,
\ee
where $i$, $j$ take values from the set $\{1;2;3\}$.
The action of the operator (\ref{Dsss1a}) on 
the lattice probability density $f({\bf m},t)$ is
\be \label{Dsss1b}
^{GL}\mathbb{D}^{\pm,\pm} 
\left[ \alpha_i \, \beta_j \atop i \ j \right] 
f ({\bf m},t) = 
\sum_{m_i=-\infty}^{+\infty} \, \sum_{m_j=-\infty}^{+\infty} \, 
K^{\pm}_{\alpha_i} (n_i-m_i) \, K^{\pm}_{\beta_j} (n_j-m_j) \, 
f ({\bf m},t) .
\ee
Equations (\ref{LattEq-2}) and (\ref{LattEq-3}) 
describe fractional
diffusion processes on the physical lattices,
where long-range jumps can be realized.
The lattice diffusion equations (\ref{LattEq-2}) and 
(\ref{LattEq-3}) can be considered
as lattice analogs of the fractional diffusion
equations for the processes with the Poissonian waiting time 
and the L\'evy distribution for the jump length.

Suggested lattice equations 
(\ref{LattEq-2}) and (\ref{LattEq-3}) 
can be considered as master equations 
that allow us to describe time-evolution of 
particles and quasi-particles on lattice since
evolution is modelled as being in exactly one of countable number of lattice sites 
at any given time, and where switching between sites is treated probabilistically. 
These equations are differential equations for the variation 
over time of the probabilities that the particle occupies each of the lattice sites.


\section{Fractional diffusion equation for continuum}


To describe fractional diffusion in the nonlocal continua,
we should use fractional derivatives with respect
to space coordinates instead of the lattice operators.
Continuum analogs of the fractional-order difference 
operators of the Gr\"unwald-Letnikov  type
are the fractional derivatives of Gr\"unwald-Letnikov type.

Fractional-order difference operators 
$^{GL}\mathbb{D}^{\pm}_L \left[ \alpha_j \atop j \right]$
defined by (\ref{Ds-GL}) 
are transformed by the continuous limit operation
into the fractional derivative
of Gr\"unwald-Letnikov type 
with respect to coordinate $x_j$ in the form 
\be 
\lim_{a_j \to 0+} \Bigl( 
\, ^{GL}\mathbb{D}^{\pm}_L \left[ \alpha_j \atop j \right] 
\, f({\bf m},t) \Bigr) =
\, ^{GL}\mathbb{D}^{\pm}_C \left[ \alpha_j \atop j \right] 
\, f({\bf r},t) ,
\ee
where 
$\, ^{GL}\mathbb{D}^{\pm}_C \left[ \alpha_j \atop j \right]$ 
are the continuum fractional derivatives
of the Gr\"unwald-Letnikov type that are defined by
\be \label{D-GLT}
\, ^{GL}\mathbb{D}^{\pm}_C \left[ \alpha \atop j \right] \, 
= \frac{1}{2} \left(
\, ^{GL}D^{\alpha}_{x_j,+} \, \pm \, ^{GL}D^{\alpha}_{x_j,-}
\right) ,
\ee
which contain the Gr\"unwald-Letnikov fractional derivatives 
$^{GL}D^{\alpha}_{x_j,\pm}$ with respect to 
space coordinate $x_j$ that can be written as 
\be \label{GLD}
^{GL}D^{\alpha}_{x_j, \pm} f({\bf r},t) = 
\lim_{a_j \to 0+} \frac{1}{|a_j|^{\alpha}} 
\sum_{m_j = 0}^{\infty} 
\frac{(-1)^{m_j} \, \Gamma(\alpha+1)}{\Gamma(m_j+1) 
\Gamma (\alpha-m_j+1)} \, f({\bf r} \mp m_j {\bf a}_j,t) ,
\quad (\alpha >0) . 
\ee
This statement can be proved by analogy with 
the proof for lattice model with 
long-range interaction of the Gr\"unwals-Letnikov 
type suggested in \cite{MOM2014}.

It is important to note that
the Gr\"unwald-Letnikov fractional derivatives 
coincide with the Marchaud fractional derivatives 
(see Section 20.3 in \cite{SKM1,SKM2})
for the functions from the space $L_r(\mathbb{R})$, 
where $1 \leqslant r <\infty$ 
(see Theorem 20.4 in \cite{SKM1,SKM2}).
Moreover both the Gr\"unwald-Letnikov and 
Marchaud derivatives have the same domain of definition.
The Marchaud fractional derivative is defined by the equation
\be 
^MD^{\alpha, \pm}_{x_j} f({\bf r},t) =
\frac{1}{ a(\alpha,s) } 
\int^{\infty}_0 \frac{ \Delta^{s,\pm}_{z_j} \, f({\bf r},t) }{
z^{\alpha+1}_j} dz_j , \quad (0<\alpha <s) ,
\ee
where $\Delta^{s,\pm}_{z_j}$ is 
the finite difference of integer order $s$,
\be
\Delta^{s,\pm}_{z_j} \, f({\bf r},t) = \sum^{s}_{k=0} 
\frac{(-1)^{k} \, s!}{(s-k)! \, k!} \, 
f({\bf r}- k \,z_j {\bf e}_j,t) ,
\ee
and $a(\alpha,s)$ is 
\be
a(\alpha,s) = \frac{s}{\alpha} \int^1_0 
\frac{(1-\xi)^{s-1}}{(\ln (1/\xi))^{\alpha}} d \xi .
\ee


Using (\ref{GL-integer}), we can note that 
the derivatives (\ref{D-GLT}) for integer orders 
$\alpha=n \in \mathbb{N}$ have the forms
\be
\, ^{GL}\mathbb{D}^{\pm}_C \left[ n \atop j \right] 
= \frac{1 \pm \, (- 1)^n}{2}
\frac{\partial^n}{\partial x^n_j} .
\ee
Therefore the continuum fractional derivatives
$\, ^{GL}\mathbb{D}^{+}_C \left[ n \atop j \right]$
are the usual derivatives 
of integer order $n$ for even values $\alpha$ only,
and the continuum operators
$\, ^{GL}\mathbb{D}^{-}_C \left[ n \atop j \right]$
are the derivatives of integer order $n$
for odd values $\alpha$ only.


For bounden lattices,
the fractional-order difference operators 
$\, _B^{GL}\mathbb{D}^{\pm}_L \left[ \alpha_j \atop j \right]$ 
defined by (\ref{Ds-GL-B}) are transformed 
by the continuous limit 
\be 
\lim_{a_j \to 0+}
\Bigl( 
\, ^{GL}_B\mathbb{D}^{\pm}_L \left[ \alpha_j \atop j \right] 
\, f({\bf m},t) \Bigr) =
\, ^{GL}_B\mathbb{D}^{\pm}_C \left[ \alpha_j \atop j \right] 
\, f({\bf r},t) ,
\ee
into the continuum fractional derivatives
of the Gr\"unwald-Letnikov type 
with respect to space coordinate $x_j$, 
\be \label{D-GLT-B}
_B^{GL}\mathbb{D}^{\pm}_C \left[ \alpha \atop j \right]= 
\frac{1}{2} \left(
\, _{x^1_j}^{GL}D^{\alpha}_{x_j,+} \, \pm \, 
_{x^2_j}^{GL}D^{\alpha}_{x_j,-} \right) ,
\ee
which contain the Gr\"unwald-Letnikov fractional operators 
defined on the finite interval $[x^1_j,x^2_j]$  
with $x^1_j=m^1_j a_j$ and $x^1_j=m^2_j a_j$,
in the form 
\be \label{GLD-B}
_B^{GL}D^{\alpha}_{x_j, \pm} f({\bf r},t) = 
\lim_{a_j \to 0+} \frac{1}{|a_j|^{\alpha}} 
\sum_{m_j = 0}^{M^{\pm}_j} 
\frac{(-1)^{m_j} \, \Gamma(\alpha+1)}{\Gamma(m_j+1) 
\Gamma (\alpha-m_j+1)} \, f({\bf r} \mp m_j {\bf a}_j ,t) , 
\ee
where 
\be
M^{+}_j = \left[ \frac{x_j-x^1_j}{a_j} \right] , \quad
M^{-}_j = \left[ \frac{x^2_j-x_j}{a_j} \right] .
\ee
The suggested forms of continuum fractional derivatives 
of the Gr\"unwald-Letnikov type allow us to consider 
diffusion processes on bounded areas of nonlocal continuum.


The lattice diffusion equation (\ref{LattEq-2}) 
in the continuum limit gives 
the fractional diffusion equation with
derivatives of non-integer orders 
with respect to space coordinates. 
This space-fractional diffusion equation
for the probability density $f({\bf r},t)$ has the form
\be \label{FFPE-1}
\frac{\partial f({\bf r},t)}{\partial t} = 
-\sum_{i=1}^3 D_i (\alpha) \, 
\, ^{GL}\mathbb{D}^{-}_C \left[ \alpha_i \atop i \right] 
f({\bf r},t) + 
\frac{1}{2} \sum_{i=1}^3 \sum_{j=1}^3 D_{ij} (\alpha,\beta) \,
\, ^{GL}\mathbb{D}^{-,-}_C \left[ \alpha_i \ \beta_j \atop i \ j\right] 
\, f ({\bf r},t) ,
\ee
where $D_i(\alpha)$ is the drift vector and 
$D_{ij}(\alpha,\beta)$ is the diffusion tensor
for the continuum that 
are defined by the lattice coupling constants
$g_{i}$ and $g_{ij}$ by the relations
$D_{i}(\alpha) = g_{i}$, 
$D_{ij} (\alpha,\beta) = 2 \, g_{ij}$.
Similarly, the diffusion equation for bounded lattice
gives the fractional diffusion equation
for bounded region of continua
\be \label{FFPE-1B}
\frac{\partial f({\bf r},t)}{\partial t} = 
-\sum_{i=1}^3 D_i (\alpha) \, 
\, ^{GL}_B\mathbb{D}^{-}_C \left[ \alpha_i \atop i \right] 
f({\bf r},t) + 
\frac{1}{2} \sum_{i=1}^3 \sum_{j=1}^3 D_{ij} (\alpha,\beta) \,
\, ^{GL}_B\mathbb{D}^{-,-}_C \left[ \alpha_i \ \beta_j \atop i \ j\right] 
\, f ({\bf r},t) .
\ee

It should be noted 
that coincidence of orders of fractional derivatives 
in the first and second terms allows us
to represent the fractional diffusion equation 
(\ref{FFPE-1}) in the form of 
the space-fractional continuity equation
\be \label{FFPE-2}
\frac{\partial f({\bf r},t)}{\partial t} = 
-\sum_{i=1}^3 
\, ^{GL}\mathbb{D}^{-}_C \left[ \alpha_i \atop i \right] 
J_i ({\bf r},t) ,
\ee
where $J_i$ is the probability flow
\be \label{PF-1}
J_i ({\bf r},t) = D_i (\alpha) \, f({\bf r},t) -
\frac{1}{2} \sum_{j=1}^3 D_{ij} (\alpha,\beta) \,
\, ^{GL}\mathbb{D}^{-}_C \left[ \beta_j \atop j \right] 
f ({\bf r},t) .
\ee
Equation (\ref{PF-1}) can be considered
as the fractional phenomenological Fick's first law
for nonlocal media.
If $\alpha_i=1$ for all $i=1,2,3$, 
the continuity equation (\ref{FFPE-2})
can be represented as the standard form of the
well-known continuity equation
\be 
\frac{\partial f({\bf r},t)}{\partial t} = 
-\sum_{i=1}^3 
\frac{\partial J_i ({\bf r},t) }{\partial x_i} ,
\ee
where $J_i ({\bf r},t)$ is defined 
by (\ref{PF-1}) with $\beta_j \ne 1$ in general.


For one-dimensional case with $D_i (\alpha)=0$ and 
$f({\bf r},t)=f(x,t)$, 
equation (\ref{FFPE-1}) can be represented in the form
\be \label{FFPE-1-1d}
\frac{\partial f(x,t)}{\partial t} = 
K(\mu) \, \nabla^{\mu} f (x,t) ,
\ee
where $K(\mu)$ is the generalized diffusion constant,
\be 
K(\mu) = \frac{1}{2} D_{11} (\alpha,\beta) , 
\ee
and $\nabla^{\mu}$ is the fractional derivative of order $\mu$, 
\be \label{nabla-1}
\nabla^{\mu} =
\, ^{GL}\mathbb{D}^{-}_C \left[ \alpha_1 \atop x \right] 
\, ^{GL}\mathbb{D}^{-}_C \left[ \beta_1 \atop x \right] =
\, ^{GL}\mathbb{D}^{-}_C \left[ \alpha_1+ \beta_1 \atop x \right] 
, \quad \mu =\alpha_1+\beta_1 .
\ee
Here we use the semi-group property of
fractional derivatives of the Gr\"unwald-Letnikov type.
Equation (\ref{FFPE-1-1d}) describes the fractional diffusion
processes with the Poissonian waiting time and 
the L\'evy distribution for the jump length
(see Section 3.5 of \cite{MK2000}).
In \cite{MK2000} the space-fractional diffusion 
equation (\ref{FFPE-1-1d}) contains 
the Weyl fractional derivative $\nabla^{\mu}$ of order $\mu$, 
of one-dimensional case.
The solution of equation (\ref{FFPE-1-1d}) can be obtained 
analytically by using the Fox function $H^{1,1}_{2,2}$
(for details see Section 3.5 of \cite{MK2000} 
and \cite{MainardiPS}).
The exact calculation of fractional moments \cite{MK2000} gives
\be \label{x-delta-2}
\langle |x(t)|^{\delta} \rangle = 
\frac{2 \, (K(\mu))^{\delta / \mu} \, \Gamma (- \delta /\mu) \, \Gamma (1+\delta)}{
\mu \, \Gamma (-\delta/2) \, \Gamma (1+\delta/2)} \, 
t^{\delta / \mu} ,
\ee
where $0< \delta <\mu \le 2$.

Using equation (\ref{FT-FD}), 
it is possible to demonstrate that
the space-fractional diffusion equations 
are connected with continuous time random walk processes 
with diverging second moment of the jump length distribution \cite{MK2000}.


If $\alpha_j=\beta_j=1$ for all $j=1,2,3$, then 
equations (\ref{FFPE-1}) and (\ref{FFPE-1B}) 
give the well-known second-order diffusion equation 
\be
\frac{\partial f({\bf r},t)}{\partial t} = -\sum_{i=1}^3 
D_i \, \frac{\partial f({\bf r},t) }{\partial x_i}+ 
\frac{1}{2}\sum_{i=1}^3 \sum_{j=1}^3 D_{ij} \,
\frac{\partial^2 f({\bf r},t) }{\partial x_i \, \partial x_j} ,
\ee
where $D_i=D_i(1)$ is the drift vector and 
$D_{ij}=D_{ij}(1,1)$ is the diffusion tensor
for local continuum.


\section{Conclusion}

Lattice analog of the fractional-order differential equations
for bounded and unbounded three-dimensional lattices with 
long-range jumps of particles are suggested.
These lattice equations can be considered
as a new microscopic basis to describe 
the fractional diffusion in nonlocal continua.
In the lattice diffusion equations, we use
the fractional-order difference analogs of 
fractional derivatives, which are represented by kernels 
that describe long-range jumps of lattice particles.
The proposed kernels of long-range jumps on the lattice 
can be considered for integer and fractional orders of 
suggested difference operators. 
The continuous limits for these diffusion equations
with fractional-order differences give 
the continuum fractional 
derivatives of the Gr\"unwald-Letnikov type 
with respect to space coordinates.
The obtained fractional diffusion on nonlocal continua
can be considered as a continuous limit of 
the suggested lattice diffusion, 
where the sizes of continuum elements are much larger than 
the distances between sites of the lattice. 
The main advantage of the suggested approach 
is a possibility to consider 
fractional-order difference diffusion equations 
as tools for formulation of a microstructural 
basic model of fractional diffusion in nonlocal continua. 
The proposed three-dimensional lattice diffusion equations 
can play an important role to formulate discrete models
of nonlocal processes in microscale and nanoscale.

It is interesting to generalize the suggested lattice approach
to consider lattice Levy flights subject 
to external force fields and the Galilean invariance. 
This transport process on lattice can be described by 
lattice fractional diffusion equations. 
We assume that the lattice Levy flights 
in a constant force field  are similar to lattice fractional diffusion in a constant velocity field 
by analogy with diffusion processes 
in continuum models \cite{JMF1999}.

We assume that the proposed lattice approach
to the lattice fractional diffusion can generalized
to different types of Bravais lattices
such as monoclinic, triclinic, hexagonal and rhombohedral
lattices.
We also assume that the suggested approach
to the fractional diffusion can be generalized for 
lattice models with the fractal spatial dispersion,
which are suggested in \cite{JPA2008} 
(see also \cite{FL-2,FL-3}), and 
the continuum limits of these fractal lattice models
can give continuum models of fractal media 
\cite{SP1985,MGN1994} that are described
by non-integer-dimensional space approach 
\cite{CNSNS2015,JMP2014}.


\end{document}